\RequirePackage[2020-02-02]{latexrelease}
\documentclass[aps, pra, preprint, groupedaddress, amsfonts,
               amsmath, amssymb, showpacs, nofootinbib]{revtex4-1}
\usepackage{microtype}
\usepackage{lmodern}
\usepackage{graphicx}
\usepackage{epsfig}
\usepackage{siunitx}
\usepackage{epstopdf}
\usepackage{changepage}
\usepackage[utf8]{inputenc}
\usepackage[T1]{fontenc}
\usepackage[usenames,dvipsnames]{xcolor}
\usepackage{hyperref}

\usepackage{color}

\begin{document}
\raggedbottom
\title{Non-monotonic dc Stark shifts in the rapidly ionizing orbitals of the water molecule}

\author{Patrik Pirkola}
\email[]{patpirko@my.yorku.ca}
\author{Marko Horbatsch}
\email[]{marko@yorku.ca}
\affiliation{Department of Physics and Astronomy, York University, Toronto, Ontario M3J 1P3, Canada}
\date{\today}
\begin{abstract}

We extend a previously developed model for the Stark resonances of the water molecule. The method
employs a partial-wave expansion
of the single-particle orbitals
using spherical harmonics. To find the resonance positions
and decay rates, we use the exterior complex scaling approach which involves the analytic continuation of the radial variable into the complex plane and yields a non-hermitian Hamiltonian matrix. 
The real part of the eigenvalues provides the resonance positions (and thus the Stark shifts), while the imaginary parts $-\Gamma/2$ are related to the decay rates
$\Gamma$, i.e., the full-widths at
half-maximum of the Breit-Wigner resonances. We focus on the three outermost (valence) orbitals, as they are dominating the ionization
process. We find that for forces directed in the three Cartesian co-ordinates, the fastest ionizing orbital always displays a non-monotonic Stark shift.  For the case
of fields along the molecular axis
we also compare results as a function
of the number of spherical harmonics included ($\ell_{\max}=3,4$). We also compare our results to the total molecular Stark shifts for the Hartree-Fock and coupled cluster methods.

\end{abstract}
%
%

\maketitle
\section{Introduction}
\label{intro}

Ionization of the water molecule by a dc electric field
been dealt with in the past by effective potential formulations, in which a local potential is formed to solve a single-electron Schr\"{o}dinger equation \cite{pm21,Laso_2017}. 
Another recent approach includes calculations using the 
Hartree-Fock (HF), and 
the correlated
coupled-cluster singles and doubles, CCSD(T) methods \cite{Jagau2018}. Molecules without external field are often treated using density functional theory \cite{Goerling2007}. Both the HF and CCSD methods are approximations to the multi-electron Schr\"{o}dinger equation \cite{Jagau2018}. In density functional theory one minimizes the total energy of the system, where one deals with the total electron density rather than the multi-electron wavefunction. 

To solve the resonance problem one employs an analytic continuation method, such as exterior complex scaling (ECS) or the inclusion of a complex absorbing potential (CAP). This is 
applied to avoid the use
of outgoing waves describing
the accelerating ionized electrons.
Analytic continuation allows the
use of square-normalizable wave
functions to describe an
exponentially decaying state.
This can be done at the level
of the $N$-electron wave function,
such as in Ref.~\cite{Jagau2018},
or at the level of molecular orbitals (MOs),
as in Ref.~\cite{pm21}.
One might criticize the
orbital-based method as 
being too restrictive by not
including electron correlation, 
but it should be noted that 
electron spectroscopy can be used to determine different first ionization energies~\cite{NixonWeb,PhysRevA.69.032701}, and therefore the
orbital picture offers 
interesting insights despite its shortcomings.
The removal of inner-shell electrons is possible by exposing molecules to X-rays and such
experiments have been carried out ~\cite{PhysRevX.11.041044}. 
Tunnel ionization for aligned molecular orbitals has been shown to provide an understanding
of harmonic generation in attosecond pulses ~\cite{Mairesse_2008}.

We have recently carried out work~\cite{pm21} to follow up on the single-center expansion HF method of Moccia \cite{Moccia64} for the field-free problem.

In the present work we apply
the partial-wave methodology to
solve for the resonance parameters for the three valence orbitals, using a finite-element implementation for the radial basis functions, which results in a matrix problem. The angular parts of the wave function are represented by complex spherical harmonics.  The spherical harmonic basis is truncated  at $\ell_{\max}=3$ for much of the
work, except for fields oriented with the molecular axis where those results
results~\cite{pm21} will be compared with
the case of $\ell_{\max}=4$. 
We use an effective potential
borrowed from the literature~\citep{ERREA201517}, and expand the hydrogenic parts in spherical harmonics, which allows for an efficient implementation. 
The main new aspect of the present
paper is the additional inclusion of the dc field along the $x$ and $y$ directions in order to observe how the Stark shifts
behave as a function of field stength. This work, therefore,
complements the findings of Jagau
for the overall Stark shift of the molecule where these directions were also investigated~\cite{Jagau2018}.

The paper is organized as follows. In Sect.~\ref{sec:model} we discuss the potential and wavefunction models employed in the calculation. In Sect.~\ref{sec:resA} we present density plots for the orbitals under the Cartesian force directions, and
in Sect.~\ref{sec:resB} the corresponding resonance parameters. Here we make our point about the non-monotonic
behavior of the dc Stark shift for the MO which ionizes most rapidly for a given field direction.
Sect.~\ref{sec:resC} provides a comparison with 
the net ionization parameters of Ref.~\cite{Jagau2018}. We offer some conclusions and an outlook in Sect.~\ref{sec:conclusions}.
Throughout the paper, atomic units (a.u.), characterized by $\hbar=m_e=e=4\pi\epsilon_0=1$, are used.

\section{Model}
\label{sec:model}
We use an effective potential for the water molecule that has been developed previously for various applications (Refs. \cite{atoms8030059,ERREA201517,PhysRevA.83.052704,PhysRevA.99.062701,PhysRevA.102.012808}). The model combines three spherically
symmetric potentials for the atoms which make up the water molecule. Each part contains a screening contribution, and the 
parameters are adjusted such that the
overall potential falls of as $-1/r$ at large distances, as is expected to avoid
contributions from electronic self-interaction.

The potential is defined as follows,
\begin{equation}\label{eq:pot0}
V_{\rm{eff}}=V_{\rm{O}}(r)+V_{\rm{H}}(r_1)+V_{\rm{H}}(r_2)\ , 
\end{equation}
\begin{align}\label{eq:pot}
\begin{split}
V_{\rm{O}}(r)=&-\frac{8-N_{\rm{O}}}{r}-\frac{N_{\rm{O}}}{r}(1+\alpha_{\rm{O}}r)\exp(-2\alpha_{\rm{O}}r)\ , \\
V_{\rm{H}}(r_j)=&-\frac{1-N_{\rm{H}}}{r_j}-\frac{N_{\rm{H}}}{r_j}(1+\alpha_{\rm{H}} r_j)\exp(-2\alpha_{\rm{H}}r_j)\ , 
\end{split}
\end{align}
where $\alpha_{\rm{O}}=1.602$, and $\alpha_{\rm{H}}=0.6170$. These determine how the screening changes as a function of the radial distance. The variables $r_j$
(with $j=1,2$) represent the electron-proton separations. The parameters defining the effective nuclear charges are given by $N_{\rm{O}}=7.185$ and $N_{\rm{H}}=(9-N_O)/2=0.9075$. The opening angle is chosen as 105 degrees, with an O-H bond length of 1.8 a.u.. These were chosen in accord with Ref. \cite{ERREA201517}.
The molecular plane is chosen as $y-z$,
and the geometric arrangement follows
the HF calculation of Moccia~\cite{Moccia64}.

The wavefunction is given as,
\begin{equation} 
\tilde{\Psi}(r, \theta, \phi) = \sum_{\ell=0}^{\ell_{\max}} \sum_{m=-\ell}^{\ell}{\sum_{i,n}^{I,N}\frac{1}{r}c_{in \ell m}f_{in} (r) Y_\ell^m(\theta,\phi) } \ , 
\label{shexp}
\end{equation}
where the $Y_l^m$ are complex-valued spherical harmonics. The functions $f_{in}$ are local basis functions on interval $i$ of the radial box. The index $n$ labels the polynomial basis functions~\cite{se93}. As outlined in Ref. \cite{pm21}, we expand the hydrogen potentials using spherical harmonics. For a basis of spherical harmonics describing the wavefunction expanded up to order $\ell_{\max}$, these potentials are expanded up to a level $\lambda_{\max}=2 \ell_{\max}$. This is validated by the selection rules imposed by the Gaunt integrals, which are expressed in terms of Wigner $3j$ coefficients.

We employ the exterior complex scaling (ECS) method, which is outlined in Refs.  \cite{PhysRevA.81.053845,se93}.
The radial variable is scaled as follows:
\begin{equation}\label{eq:rs}
r \to r_{\rm s}+(r- r_{\rm s})e^{i \xi}.
\end{equation} 
This scaling is applied to the radial variable wherever it appears in the Hamiltonian (cf. Ref~\cite{pm21}).

To extend the previous work to fields
in more directions than along the
molecular axis ($\hat z$) we can
write the Schr\"{o}dinger equation, e.g.,
for a water molecule with a force experienced by the electron, with behavior ${\vec F} \sim \hat x$ as
\begin{equation}\label{eq:schro1}
\bigg[\frac{-1}{2}\nabla^2 - \sum_{i=1}^{3}\frac{Z_i(|\vec{r}_i|)}{|\vec{r}_i| } - F_x r  \sin(\theta)\cos(\phi)\bigg]\Psi = E\Psi.
\end{equation}
The radial variable $r \equiv r_3$ (which appears
also in the expanded hydrogenic $|\vec{r}_i|$-dependent parts for $i=1,2$) is scaled in
accord with Eq. (\ref{eq:rs}).
The screening functions $Z_i(r)$
are obtained from Eqs.~(\ref{eq:pot})
by multiplication with the appropriate
radial coordinate.
A description of the ECS methodology can be found in Refs. \cite{pm21,PhysRevA.81.053845,se93}.

In addition to fields oriented along the $x$-axis and the  $y$-axis, we also implemented fields oriented in the $x-z$ plane (the results for which are given in the Supplementary Material [link to be provided]). 
Their representation in terms
of spherical harmonics is straightforward
as indicated in the Supplementary Material ([link to be provided]).
We focus in this paper on the field orientations along the Cartesian axes, since 
they result in an interesting behavior of the dc Stark shift for the MO which is most effectively ionized
for a given field direction.
We present the data as a function of the force as experienced by the electron, rather than the field.

For the scaling radius we have chosen a large value, 16.4 a.u. (compared to the molecular size) so that we can scale the potential of the molecule when it is approximated well by $-1/r$. The scaling angle is approximately 1.4 radians, and the radial functions extend to $r_{\max}=24.3$ a.u.. The FEM implements
Neumann conditions at $r_{\max}$, and we have enforced a boundary condition of $\Psi=0$ at $r=0$.

\section{Results}\label{sec:res}

\subsection{Visualization of molecular orbitals for $x$-, $y$-, and $z$-directed forces}\label{sec:resA}

We have calculated the orbitals from the ECS-FEM eigenvectors, rather than the CAP-FEM solutions from
Mathematica\textsuperscript{\textregistered}  presented in Ref. \cite{pm21}. The shapes are generally consistent between these methods to the point that visual inspection alone cannot identify differences (the color schemes are similar, but not identical).
For the same force orientation in Ref. \cite{pm21} we now add a comparison of $\ell_{\max}$ 3 and 4.

\begin{figure*}[h]
\centering

\includegraphics[width=300pt]{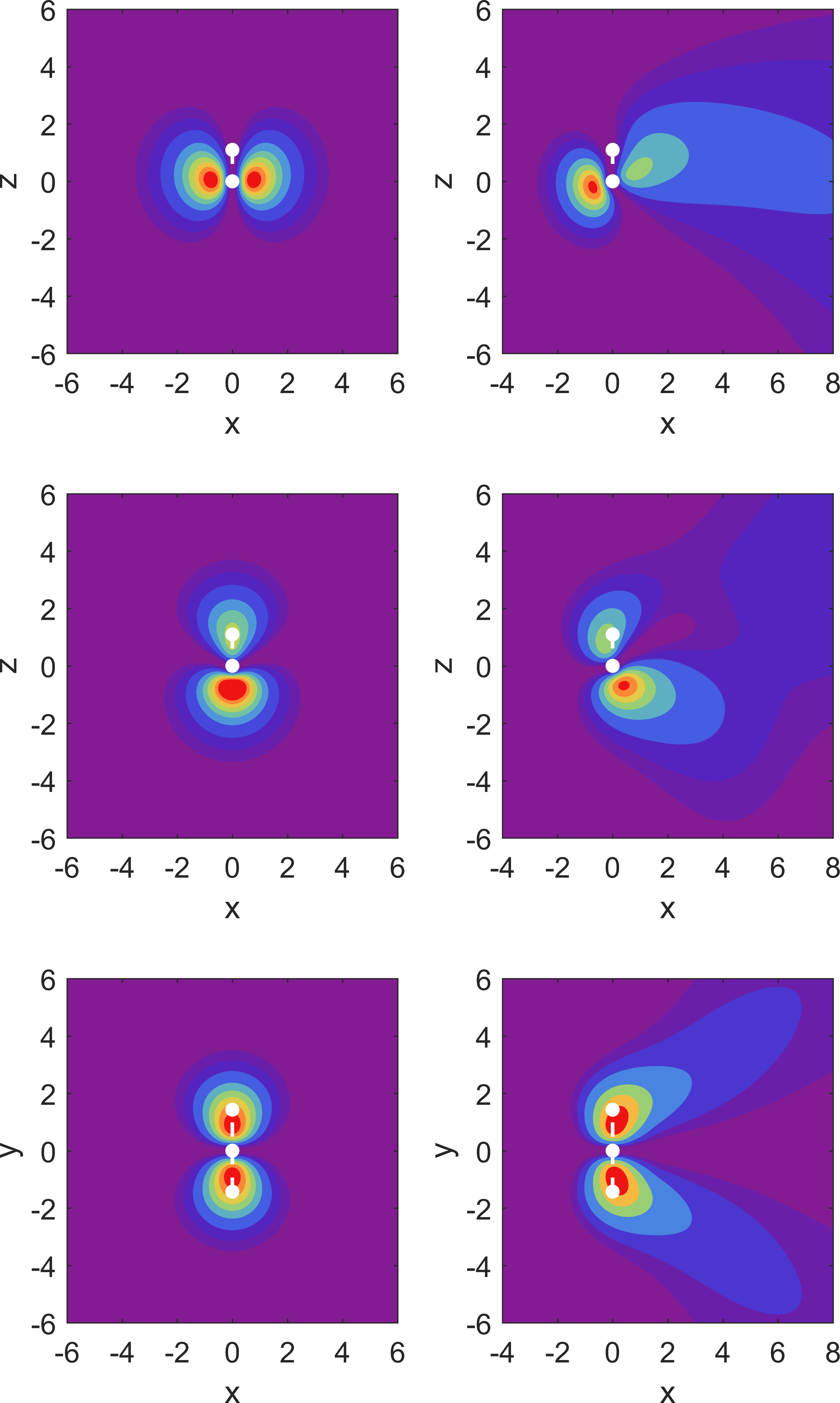}
\caption{\label{fig:denx}
Probability densities of the MOs calculated at the level of $\ell_{\max} =3$ with an $x$-directed force:  $1b_1$ (top row), $3a_1$ (middle row) and $1b_2$ (bottom row). On the left are the field-free cases, 
and on the right with electric force applied with strengths 0.12, 0.20, 0.28 a.u. respectively, which give similar decay rates. Contour heights starting with the outermost are 0.005, 0.01, 0.02, then they increase in steps of 0.02 a.u.. The positions of the nuclei are marked by white dots.}
\end{figure*}

All density plots show some peculiarities. For the force applied in the $x$ direction, the $1b_1$ MO in Fig. \ref{fig:denx} shows a structure in the orbital shape, where the left lobe of the orbital bends down, along with the right lobe of the orbital. In other words, the electron density is not purely directed along the direction of the force. For the $3a_1$ MO we observe that it is pushed toward the top right of the box, which implies that the $x$ directed force leads to part of the density being pushed in the positive $z$ direction. For the  $1b_2$ MO there is a run-away of the density to both positive and negative $z$ along with a movement towards positive $x$. 

\begin{figure*}[h]
\centering

\includegraphics[width=300pt]{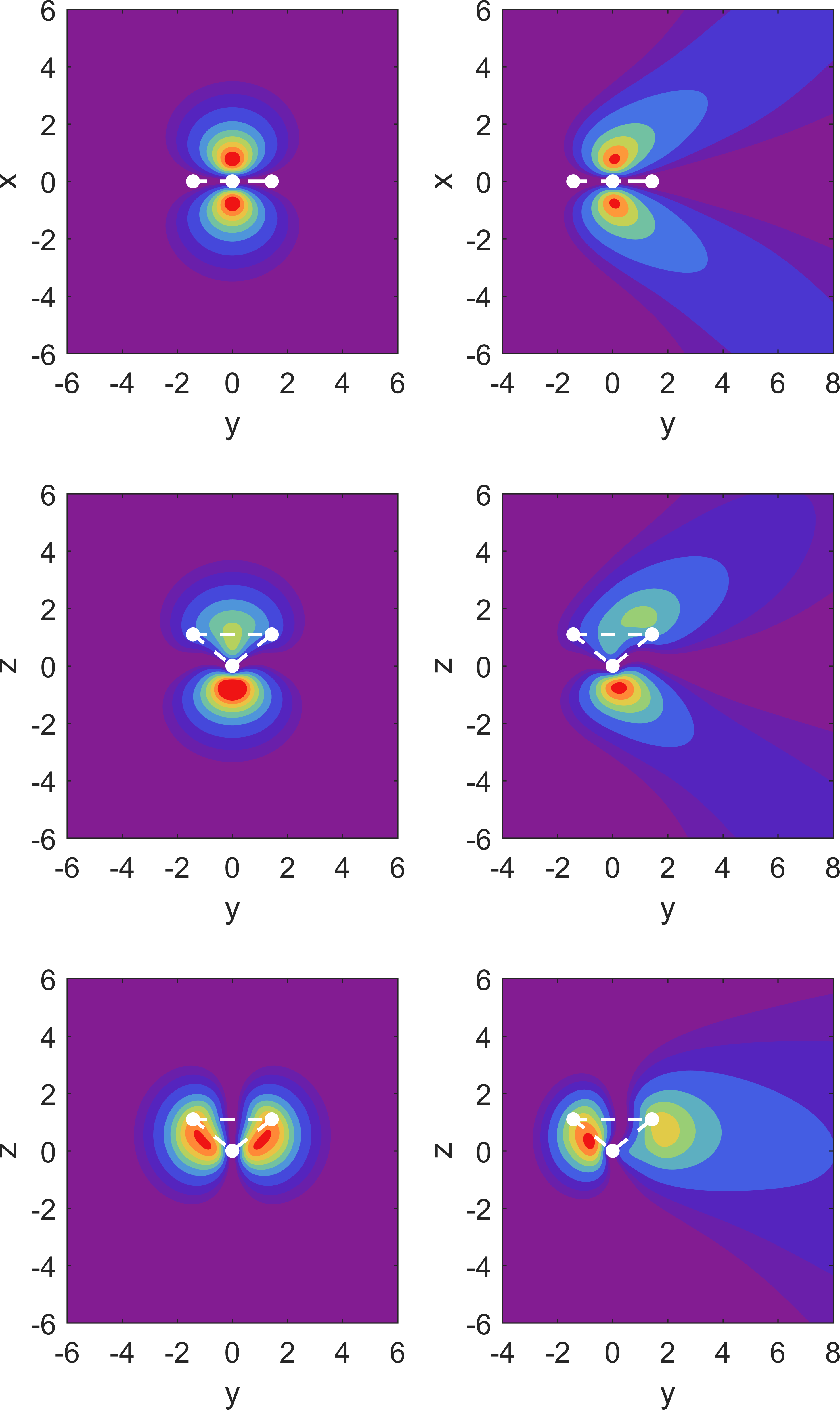}
\caption{\label{fig:deny}
Probability densities of the MOs calculated at the level of $\ell_{\max} =3$ with a $y$-directed force: $1b_1$ (top row), $3a_1$ (middle row), and $1b_2$ (bottom row). On the left are the field-free cases, and on the right with electric force applied with strengths 0.20, 0.20, 0.16 a.u. respectively, which give similar decay rates. Contour heights starting with the outermost are 0.005, 0.01, 0.02, then 
they increase in steps of 0.02 a.u.. The positions of the nuclei are marked by white dots.}
\end{figure*}

While looking at results for the $y$-directed force in Fig.~\ref{fig:deny} one notices the continuation of a 
trend  as the MOs $1b_1$ and $3a_1$ also have a run-away movement of the density away from the direction of the force. Finally, for $1b_2$ we also observe a structure in the right lobe downwards, away from the direction of the force. 
All these deformations of the MO probability densities 
reflect the combined influence of molecular potential and external dc electric field potential.

\begin{figure*}[h]
\centering

\includegraphics[width=300pt]{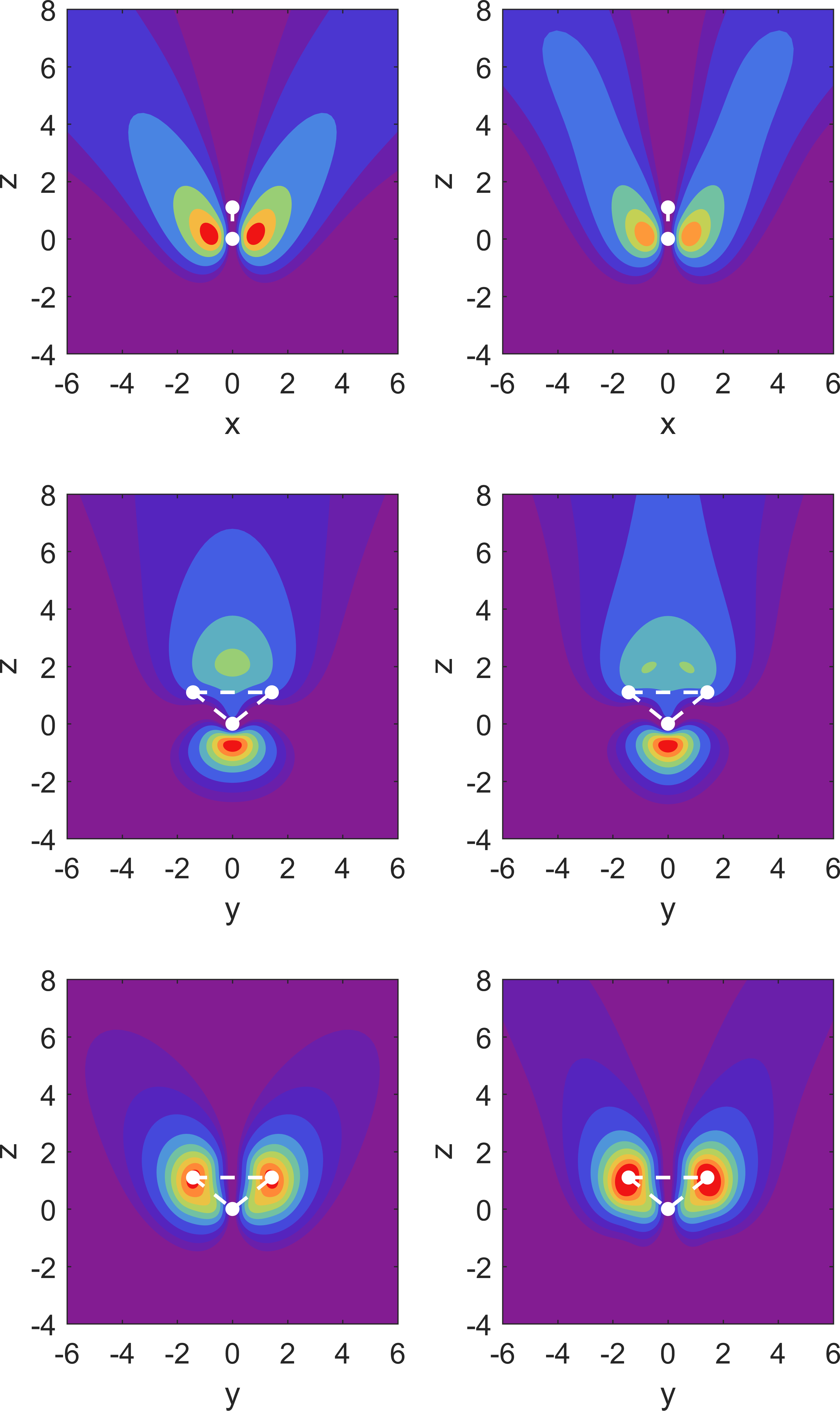}
\caption{\label{fig:denlmax}
Probability densities using a $z$ directed force for the MOs $1b_1$ (top row), $3a_1$ (middle row) and $1b_2$ (bottom row).
Left column: based on $\ell_{\max}=3$, right column for $\ell_{\max}=4$.
The force strength for the MO $3a_1$ is 0.10 a.u., while for $1b_1$ and $1b_2$ it is 0.20 a.u.. 
Contour heights starting with the outermost are 0.005, 0.01, 0.02, then they increase in steps of 0.02 a.u..}
\end{figure*}
\clearpage

\label{sec:DCmo}

In Fig. \ref{fig:denlmax} we display the MO densities for the case when 
the force is directed in the $z$ direction. Here we carry out a comparison between $\ell_{\max}$ 3 and 4. When increasing $\ell_{\max}$ we observe: {\it (i)} a narrowing and re-direction of the MO density of $1b_1$ towards the direction of the force; {\it (ii)} the same occurs for $3a_1$, although there are also two bumps in density which form at the base of the uppermost lobe; and {\it (iii)} we find that the MO $1b_2$ slightly changes in shape, and the lobes lengthen, in part towards the direction of the force. This is caused by the added flexibility in the partial-wave expansion.

To summarize the visualization of our results, which required added computational effort compared to the resonance parameter calculations, we note that the spatial distribution of ionized electrons even under dc field conditions is far from trivial, and would present a fertile ground for comparison with experiment. We note that in the context of ac laser field ionization R-matrix theory was used recently to predict interesting ionization patterns for the water molecule~\cite{PhysRevA.102.052826}.
\subsection{Dominant field direction for ionization of MOs and their dc Stark shift}\label{sec:resB}
\label{sec:DCmo}

We now proceed with a presentation of the resonance parameters for the MOs, and focus on the behavior of the
dc Stark shift for the MO with the largest ionization rate for a given field direction.
Our main observation in this work is that the orbital with the strongest ionization rate for a given external field direction displays a non-monotonic behavior in the dc Stark shift.

\begin{figure*}[h]
\centering
\hspace*{-1.4cm}
\vspace*{-0.25cm}
\includegraphics[trim=0cm 0cm 0cm 1.8cm, width=550pt]{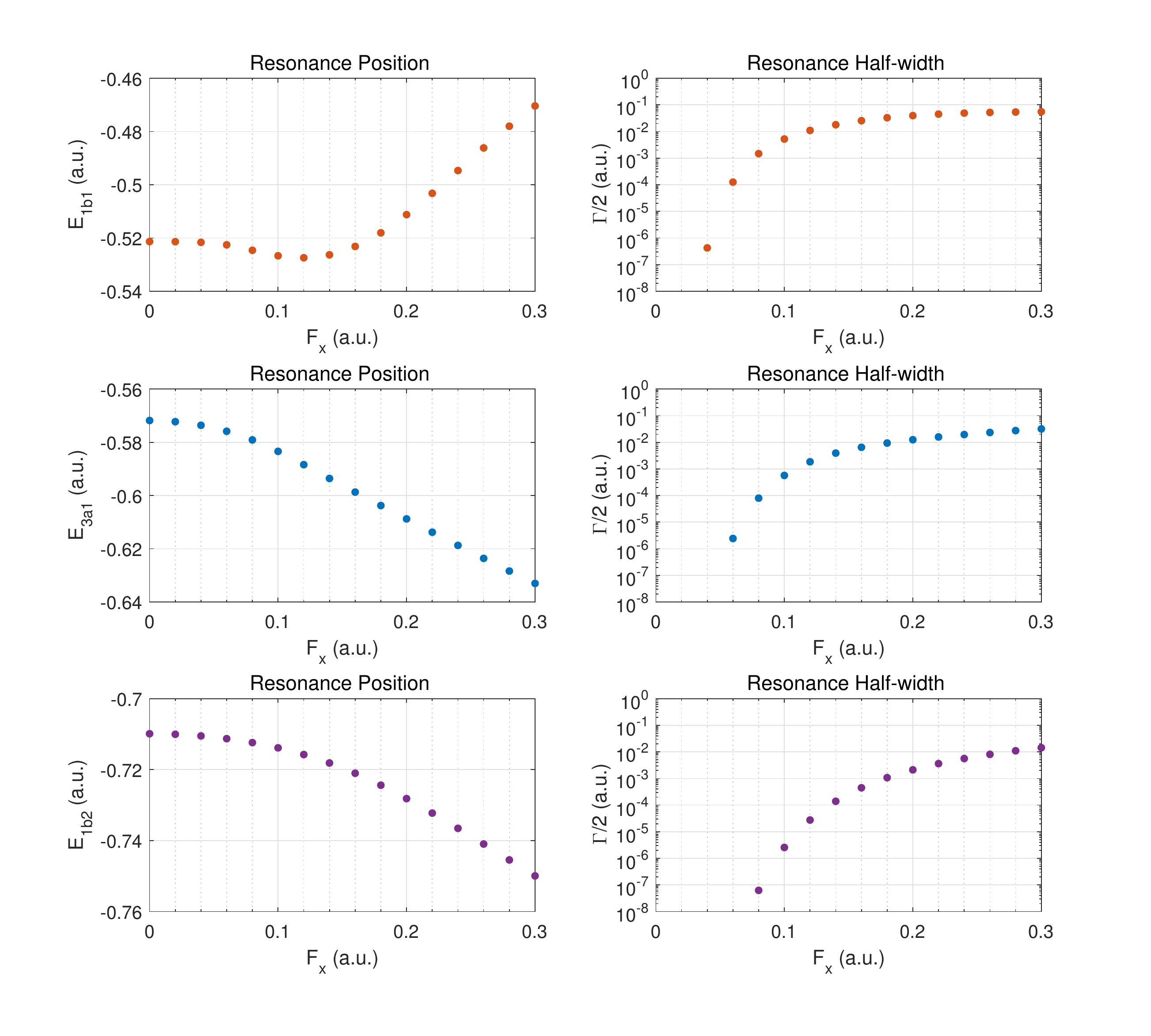}
\caption{\label{fig:xfields}
The left panel contains the resonance positions while the right panel contains half-widths for the valence orbitals of water in a DC field. Here $F_x > 0$ corresponds to a force experienced by the electron in the positive $x$ direction.}
\end{figure*}

Graphs of the resonance position for fields along the
molecular axis $\hat z$
as a function of field strength were shown
in Ref.~\cite{pm21}  to display such behavior
for the $1b_1$ orbital for the orientation
where electrons are pushed out opposite to the hydrogen atoms. The shift was initially positive,
then reached a maximum, became zero at about
$F_z=0.14$ a.u., in order to continue to negative values for stronger fields. 

For the $3a_1$ orbital non-monotonic dependence
of the the dc shift was observed in both directions with minima occurring at force
strength of the order of $F_z=-0.22$ a.u. and 
$F_z=0.14$ a.u. respectively. 
The $3a_1$ MO is the orbital with the strongest ionization rate in that case,
and the minimum in the shift at $F_z=0.14$ a.u.
corresponds to an almost vanishing dc Stark shift.
The bonding $1b_2$ MO
on the other hand showed monotonic behavior for fields along $\hat z$.
The purpose of this section is to show that these features can be generalized to
other field directions.

In Fig.~\ref{fig:xfields} the three valence
orbitals are shown for the case of an external
field perpendicular to the molecular plane.
The results are obviously symmetric with respect to reversing the field orientation.
The half-widths are showing clearly that the
MO oriented with lobes in this direction ($1b_1$)
is the most easily ionized valence orbital with rates exceeding the
other MOs by one or more orders of magnitude before saturation sets in.

While the $1b_1$ MO shows non-monotonicity in the dc Stark shift with a minimum at $F_x=0.12$ a.u.,
and a vanishing shift at $F_x \approx 0.16$ a.u. the
other MOs simply acquire an increasingly negative
dc shift.

We can now ask whether this behavior is more than coincidental: is it true that
for a field oriented along the $\hat y$ direction the bonding $1b_2$ orbital might be
affected similarly to the $1b_1$ MO's behavior in the case of a field along $\hat x$?

The answer is provided in Fig.~\ref{fig:yfields} below:
Indeed, despite its deep binding energy for substantial field strengths, such as $F_y>0.1$ a.u.
this orbital is clearly the most easily ionized
of the three valence orbitals. Note that this is
not the case for weak fields, i.e., in the pure
tunneling regime.

The reason for the dominance as far as ionization rate is concerned, geometric considerations are, of course, an important reason: in a simplistic representation of the three orbitals, i.e., $1b_1 \approx 2p_x$, $1b_2 \approx 2p_y$, and $3a_1 \approx 2p_z$ it is obvious that they respond strongly to fields aligned with $x,y,z$ respectively due to the occurrence of 
substantial dipole matrix elements from the external field.

Interestingly, the non-monotonic behavior in the dc Stark shift for the $1b_2$ orbital occurs only for field strength $F_y>0.1$ a.u., with a minimum
at $F_y\approx 0.18$ a.u., and a vanishing shift
at $F_y \approx 0.25$ a.u.. Thus the overall trend is best comparable to the orbital $1b_1$ which displays very similar behaviour when under a force in the $x$ direction. The trends of both these orbitals under their respective forces outside of $F > 0.1$ a.u. are very similar to the trend in $3a_1$ when the $z$ force is in the negative direction.

\begin{figure*}[!h]
\centering
\hspace*{-1.4cm}
\vspace*{-1.25cm}
\includegraphics[trim=0cm 0cm 0cm 0.6cm, width=550pt]{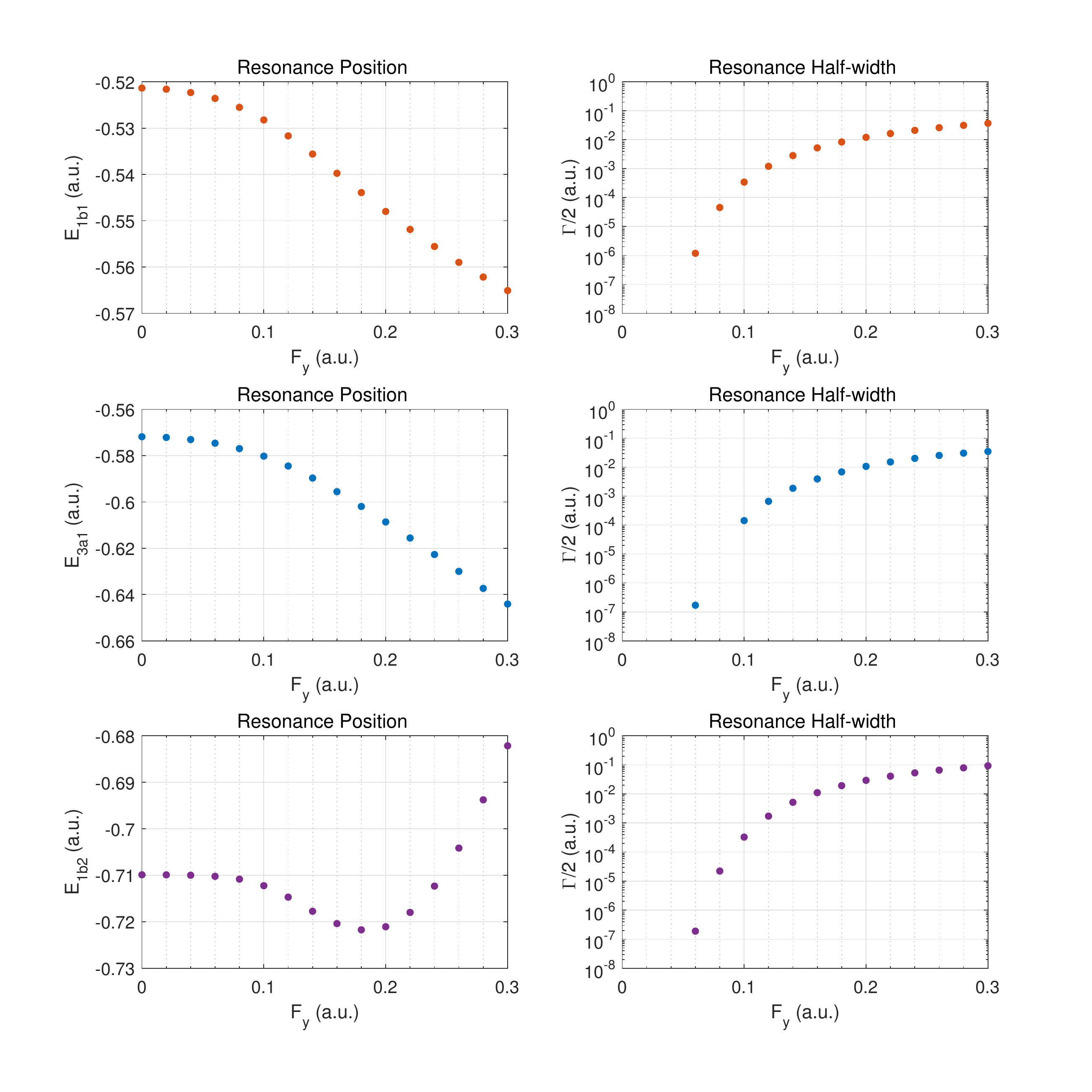}
\caption{\label{fig:yfields}
The left panel contains the resonance positions while the right panel contains half-widths. Here $F_y > 0$ corresponds to a force experienced by the electron in the positive $y$ direction.}
\end{figure*}
\clearpage

\begin{figure*}[h]
\centering
\hspace*{-1.4cm}
\vspace*{-1.25cm}
\includegraphics[trim=0cm 0cm 0cm 0cm, width=550pt]{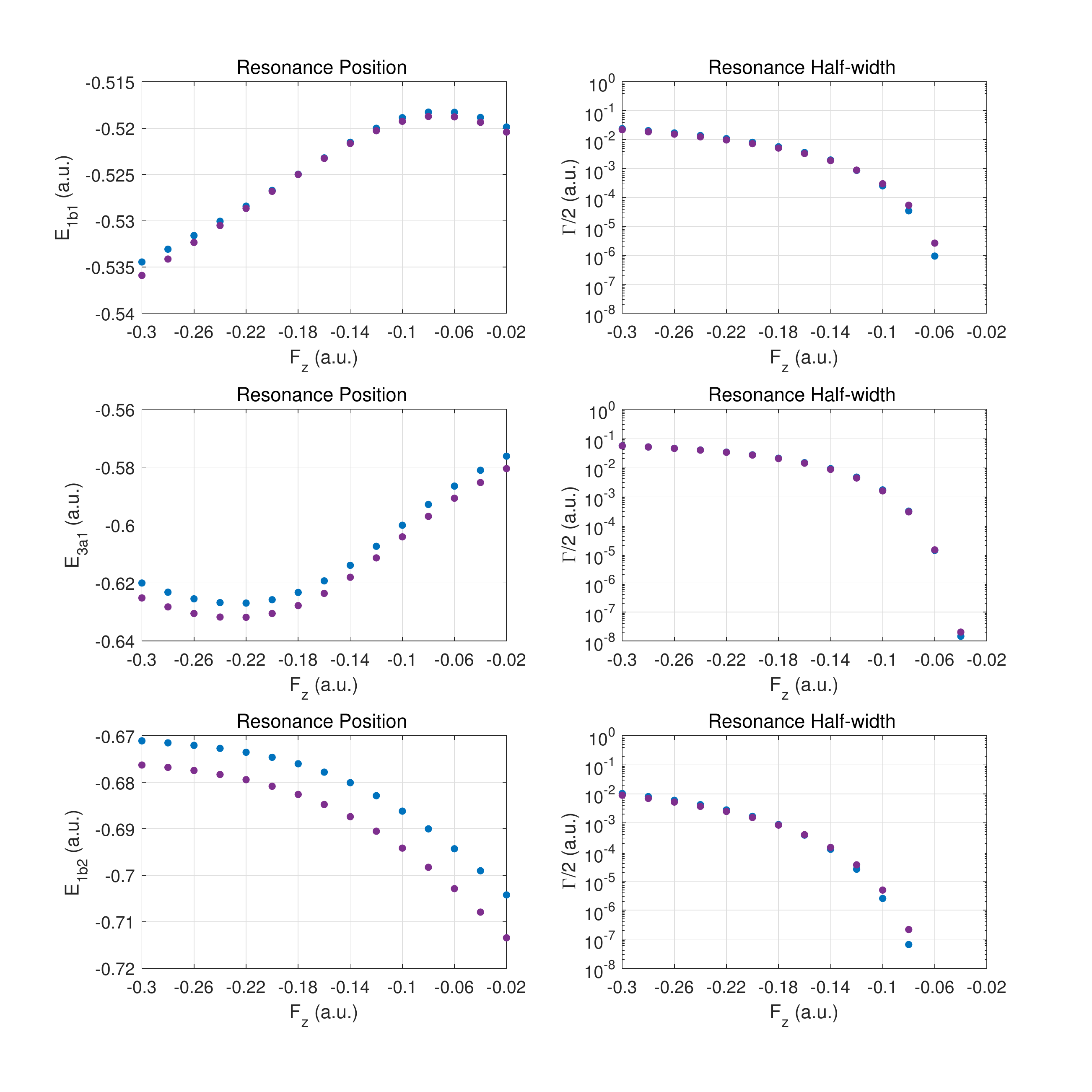}
\caption{\label{fig:lmax4}
The left panel contains the resonance positions while the right panel contains half-widths for the valence orbitals of water in a DC field. Here $F_z < 0$ corresponds to a force experienced by the electron in the negative $z$ direction. The $\ell_{\max}=3$ values are listed by blue $\textcolor{NavyBlue} \bullet$ (ECS). The $\ell_{\max}=4$ values are listed by purple $\textcolor{Purple} \bullet$ (ECS). }
\end{figure*}

\begin{figure*}[h]
\centering
\hspace*{-1.4cm}
\vspace*{-1.25cm}
\includegraphics[trim=0cm 0cm 0cm 1.6cm, width=550pt]{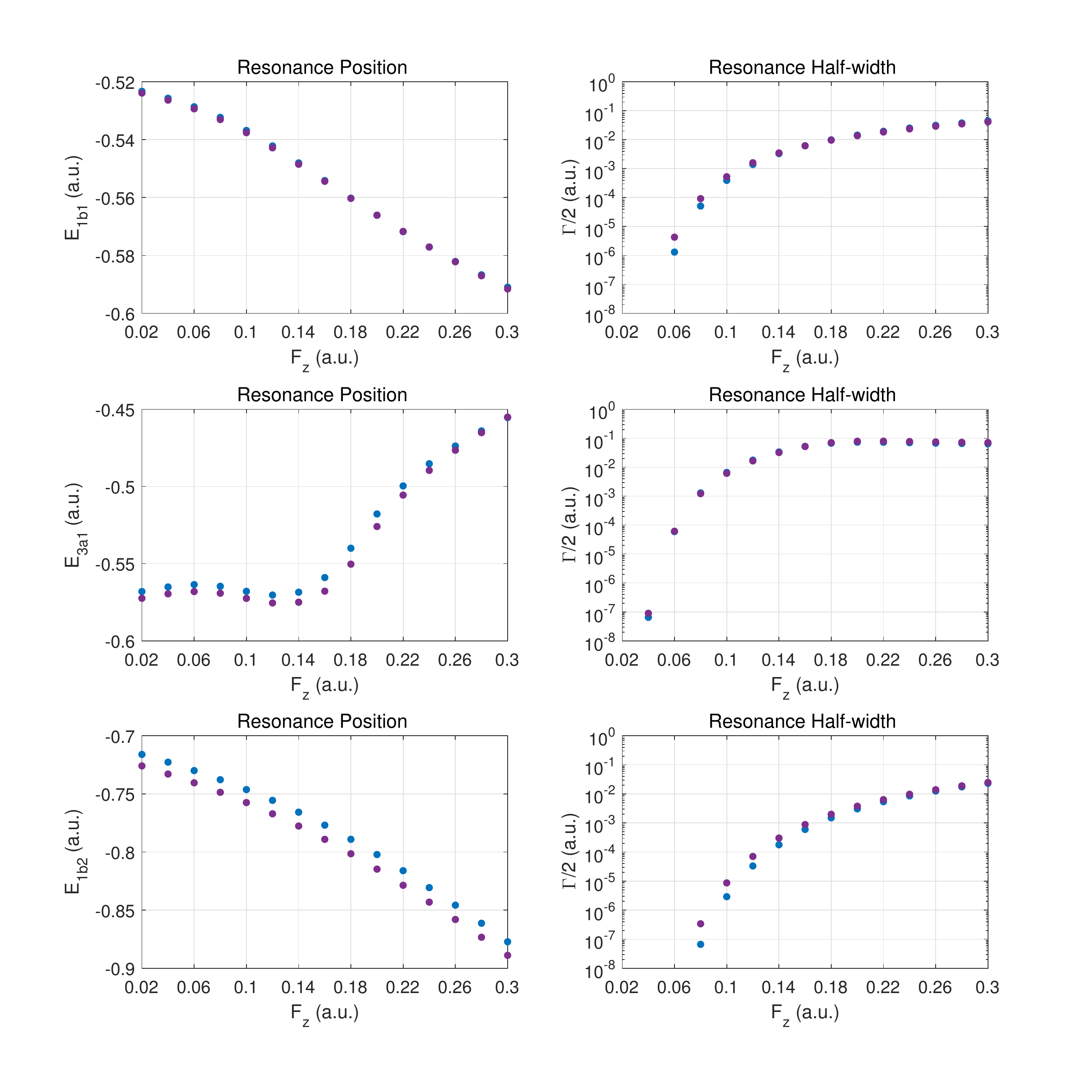}
\caption{\label{fig:lmax4_2}
The left panel contains the resonance positions while the right panel contains half-widths for the valence orbitals of water in a dc field. Here $F_z > 0$ corresponds to a force experienced by the electron in the positive $z$ direction. The $\ell_{\max}=3$ values are listed by blue $\textcolor{NavyBlue} \bullet$ (ECS). The $\ell_{\max}=4$ values are listed by purple $\textcolor{Purple} \bullet$ (ECS).}
\end{figure*}

For the case of fields along the molecular axis $\hat z$ which was discussed in Ref.~\cite{pm21}, we report some additional results here, having extended the calculations to the level of $\ell_{\max}=4$. 
Compared to the field orientations along $\hat x$ and $\hat y$ we have
the complication of asymmetry of the molecule along the molecular axis, and, thus, we separate the presentation for fields aligned such that the 
force pushes electrons out along this axis
(Fig.~\ref{fig:lmax4}), and in the opposite
direction (Fig.~\ref{fig:lmax4_2}).

We begin with electrons being pushed out towards the oxygen atom. When one considers field strengths beyond the tunneling regime (where the width turns over towards saturation) the $3a_1$ MO emerges as the one with the highest ionization rate.
Due to the asymmetry the dc Stark shift has a non-zero slope at zero field. The resonance position decreases with increasing field and reaches a minimum at $F_z\approx 0.23$ a.u..
The higher-convergence $\ell_{\max}=4$ results run
parallel to those for $\ell_{\max}=3$ for
the dc shift. For the resonance width the two results are not well distinguishable on a logarithmic scale except for the weakest field strengths shown.

Concerning the convergence with $\ell_{\max}$ we can state that the weakest bound MO ($1b_1$) is showing the smallest discrepancy, while the bonding orbital ($1b_2$) is affected most, because the partial-wave expansion is not yet sensitive to the full potential from the hydrogenic parts.
We observe, however, that the convergence (or lack thereof) does not have a strong impact on the features reported in this work, i.e., the non-monotonicity of the shifts.

For fields in the opposite direction, i.e., ionization into the half-space on the hydrogenic side, we again notice that the dominant ionization contribution is from the MO $3a_1$, and that this MO displays complicated non-monotonic behavior in the dc Stark shift.

To summarize this section we observe that the conclusions are consistent for the four possible field orientations (the $z$ direction has two possible orientations due to asymmetry of the molecule) associated with symmetry axes of the orbitals. To phrase it
simply: the non-monotonic behavior of the dc Stark shift goes hand in hand with the relatively strong ionization rate for the orbital of that particular symmetry. 

While our results for the widths of the $3a_1$ orbital
show an asymmetry when changing the $z$ force direction from positive to negative,
this behavior is different from what was observed for the net molecular width
in HF theory~\cite{Jagau2018}, as shown in Fig.~7 of Ref.~\cite{pm21}.

\clearpage

 Our own
net widths are dominated by the $3a_1$ orbital, and so a natural question to ask is 
whether there is a convergence issue in our partial-wave approach. The $\ell_{\max}=4$
results do not deviate, however, substantially from those for $\ell_{\max}=3$.
Thus, one will have to investigate further whether this difference in behavior is
related to the determination of the net decay width, or whether it is the model
potential approach that fails to account for self-consistent field effects in
the presence of the external field.

\subsection{Comparison with HF and CCSD(T) calculations for net ionization}\label{sec:resC}


\begin{figure*}[!h]
\centering
\hspace*{-1.65cm}
\includegraphics[width=550pt]{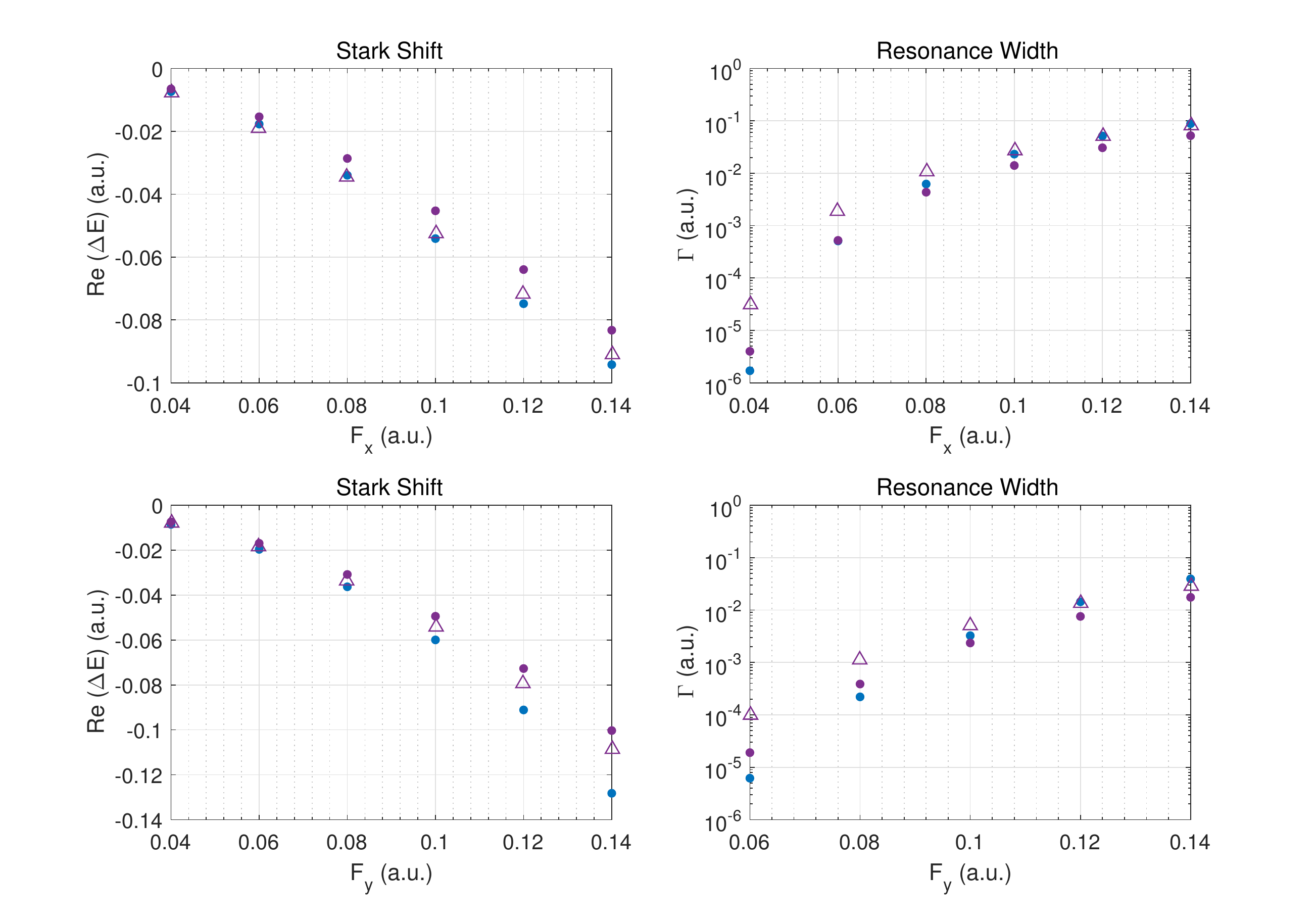}
\caption{\label{fig:jagau2}
The left panel displays dc Stark shifts, while the right panel displays full-widths. The blue $\textcolor{NavyBlue} \bullet$ are the present DS (2) results calculated at the
level of $\ell_{\max}=3$, the purple $\textcolor{Purple} \bullet$ are the HF results, and the purple $\textcolor{Purple} \triangle$ are the CCSD(T) results.}
\end{figure*}
\clearpage

In Fig.~\ref{fig:jagau2} we compare with the HF and CCSD(T) total molecular Stark shifts and resonance widths given in Ref. \cite{Jagau2018}. 
As explained in Ref.~\cite{pm21} for our model 
calculation a meaningful
comparison is to consider the direct sum of orbital
energies. Such an
analysis corresponds to total, (or net) ionization from the molecule, and the five MOs are counted
twice to account for the spin degeneracy.

The most direct comparison for the
present results should be with the mean-field
single-particle approach, i.e., the HF method, as our model potential 
is designed to match HF orbital energies.
The comparison of the dc Stark shifts for field
directions along $\hat x$ and $\hat y$ shows that
the present model calculation yields stronger
shifts as the field strength increases, but that
the overall trend agrees with the HF results of Ref.~\cite{Jagau2018}. 

We observe a similar trend in the decay widths:
they agree at the factor-of-two level
for the cases involving fields along $\hat x$ and $\hat y$. 
This is an improvement
compared to the results for forces along the
molecular axis in the positive $z$ direction, while comparable to the
case when the force is in the negative $z$ direction \cite{pm21}. We note that the additional
corrections due to electronic correlation,
i.e., the CCSD(T) over the HF results is also
on a similar scale, i.e., a discrepancy at the level of factors of two-three
when the ionization rate is strong.

The blue $\textcolor{NavyBlue} \bullet$ mark the DS (2) values, i.e., the direct sum method, in which we calculate the total molecular Stark shift and width by adding the resonance parameters for every MO (as reported in Sect.~\ref{sec:resB}) assuming double occupancy due to spin degeneracy~Ref.~\cite{pm21}. The agreement between DS (2) and the HF results of Ref.~\cite{Jagau2018}
is generally good for the Stark shifts but less so for the widths. 
At higher field strengths, the agreement weakens, and more so for the force in the $y$ direction. 
For the widths, DS (2) in $\textcolor{NavyBlue} \bullet$ begins with improving in agreement with HF in $\textcolor{Purple} \bullet$ relative to the the force strength. We note that the correlated CCSD(T) calculations of Ref.~\cite{Jagau2018} generally do not deviate much from the HF data, and that
the present model calculations in some cases, perhaps fortuitously, agree with them.

The Stark shifts for the $x$-directed force, agree very well with the CCSD(T) results of Ref.~\cite{Jagau2018} with single-digit percentage deviations. 
It would be of interest to compare the MO resonance parameters from HF calculations
with the present model potential results in order to complement the comparison
of net quantities which follow from the total energy.
For the purposes of comparison with experiment, and understanding the theory more clearly, more work must be done to bridge the gap between the multi-electron solutions, HF and CCSD(T), and the present single-electron, local potential approach.
For future work it is planned to extend the current work to a potential model 
from density functional theory, such as the local HF potential method~\cite{SalaGoerling2002,Sala2007}.

\section{Conclusions}\label{sec:conclusions}
In this work we have extended our previous model calculations for dc field ionization of the water molecule~\cite{pm21}  mostly in two respects:

{\it (i)} we have included two orientations for
the external field to complement the previous
work which was restricted to fields along the
molecular axis;

{\it (ii)} for fields oriented with the molecular axis
we have validated  the conclusions based on the limited calculations in the 
angular momentum basis by comparing results for $\ell_{\max}=3$
with those for  $\ell_{\max}=4$. 

Extension {\it (i)} allowed us to gain some understanding concerning the non-monotonic 
behaviour of the dc Stark shift, as being
associated with the MO that is most easily ionized by a given field orientation.
Extension {\it (ii)}, while not a complete convergence analysis, nevertheless provides strong evidence that the
major conclusions concerning dc Stark shifts and
resonance widths will not be overturned by the inclusion of more partial waves. These higher-$\ell$ contributions are likely to play a significant role when one analyzes the spatial emission properties of the ionized electrons.

It would be of great interest to explore in experiments with infrared laser fields and oriented molecules 
the predictions made for the non-monotonic dc Stark shifts. Experiments with oriented nitrogen and carbon
dioxide molecules have been performed~\cite{Mairesse_2008}, and water vapor does represent a challenge.
Ionization from particular MOs would require some method of vacancy detection, so this is definitely a 
challenge compared to ionization from particular MOs by electron~\cite{ElectronsWater2021} or X ray~\cite{PhysRevA.102.052826} impact where one has some control
through the incident particle energy, or even the secondary electron energy in an (e,2e) process~\cite{NixonWeb}.

\begin{acknowledgments}
Discussions with Tom Kirchner and Michael Haslam are gratefully acknowledged. We would also like to thank Steven Chen for support with the high performance computing server used for our calculations.
Financial support from the Natural Sciences and Engineering Research Council of Canada is gratefully acknowledged. \end{acknowledgments}

\bibliography{Partial2}
\end{document}


\raggedbottom
\title{Non-monotonic dc Stark shifts in the rapidly ionizing orbitals of the water molecule (Supplementary Materials)}

\author{Patrik Pirkola}
\email[]{patpirko@my.yorku.ca}
\author{Marko Horbatsch}
\email[]{marko@yorku.ca}
\affiliation{Department of Physics and Astronomy, York University, Toronto, Ontario M3J 1P3, Canada}
\date{\today}
\maketitle

\section{Spherical harmonic decomposition of the external field}
The potential for a force oriented in the $x$ direction can be broken down using spherical harmonics. First, one knows that complex exponentials can be used to express the spherical-polar form of the potential. We write this as,
\begin{align}\label{eq:shfield}
\begin{split}
- \frac{F_x r}{2}  \bigg(\sin(\theta)\exp(i\phi)+\sin(\theta)\exp(-i\phi)\bigg) \\= - \frac{F_x r \sin(\theta)}{2}  \bigg(\cos(\phi)+i\sin(\phi)+\cos(\phi)-i\sin(\phi)\bigg)
\\= - F_x r \sin(\theta)  \cos(\phi).
\end{split}
\end{align}
This will be important for the spherical part of the integrals in the matrix elements, as the first line in Eq. (\ref{eq:shfield}) contains a pair of spherical harmonics, but care must be taken of normalization, since we use Wigner $3j$ coefficients to evaluate the Gaunt integrals which are involved in the matrix elements. The Wigner coefficients are defined for the normalized spherical harmonics involved in the integrals, and thus to retain the result above in Eq. (\ref{eq:shfield}) the result from the Wigner coefficients must be multiplied by the appropriate coefficient. A little more clearly, we shall split the integral involved for the DC potential as follows,
\begin{align}\label{eq:shfield2}
\begin{split}
- \frac{F_x }{2} \big\langle inlm\big| r\sin\theta\exp i\phi \big| i'n'l'm'\big\rangle - \frac{F_x }{2} \big\langle inlm \big| r\sin\theta\exp-i\phi \big| i'n'l'm'\big\rangle
\\=- \frac{F_x }{2} \big\langle in \big| r \big| i'n'\big\rangle\bigg(\big[ lm \big| \sin\theta \exp  i\phi \big| l'm' \big] + \big[  lm \big| \sin\theta \exp-i\phi \big| l'm' \big]\bigg).
\end{split}
\end{align}
In the first line we have used standard angled Dirac brackets. However, in the second line we have split the matrix elements up into two sets of brackets. In the second line the first angled set is similar to Dirac brackets and indicate integration, but only over the radial variables and radial basis functions, and the square brackets indicate the same but for the spherical parts. The results of these square brackets will be replaced by Wigner $3j$ coefficients for the corresponding spherical harmonics multiplied by the appropriate constants, which are $2\sqrt{2\pi/3}$ and $-2\sqrt{2\pi/3}$ for $(l,m)$ equal to $(1,-1)$, and $(1,1)$ respectively. These constants are just the multiplicative inverse of the spherical harmonic normalization coefficients.
\section{Field orientation in the $x-z$ plane: resonance parameters for $1b_1$ and $3a_1$}
\begin{figure*}[!h]
\centering
\hspace*{-1.65cm}
\includegraphics[width=550pt]{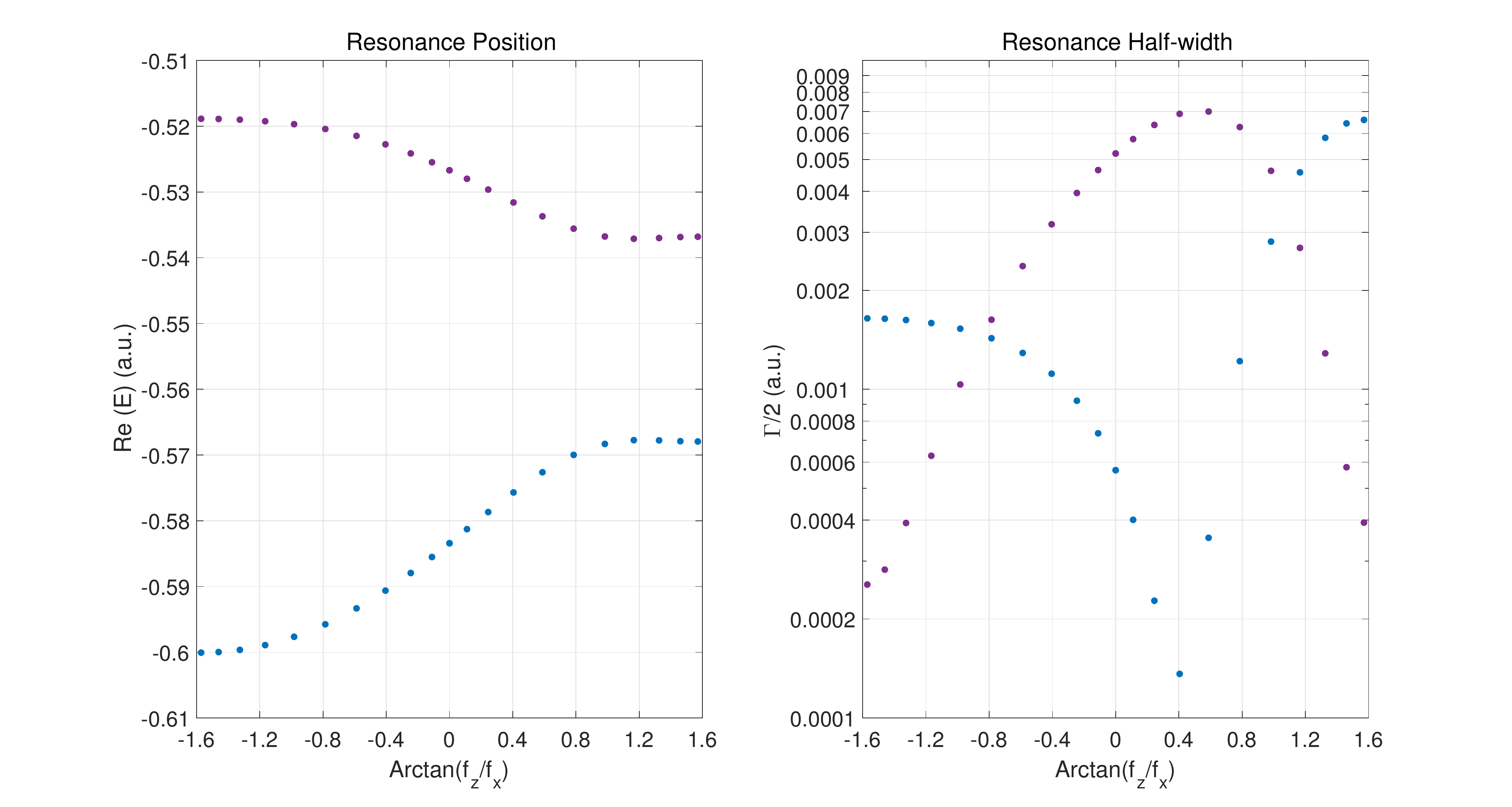}
\caption{\label{fig:both}
The left panel contains the resonance positions while the right panel contains half-widths for the $3a_1$ (blue $\textcolor{NavyBlue} \bullet$) and $1b_1$ (purple $\textcolor{Purple} \bullet$) valence orbitals of water in a dc field ($F_x > 0$). The force strengths in each direction are $F_z = F_of_z$ and $F_x = F_o f_x$, where the lower-case $\{f_x, f_z \}$ are the dimensionless fractions which determine the projection of $\vec{F}$ onto the two axes. Hence the angle of the force from the $x$ to the $z$ axis is $\arctan{(F_of_z/F_of_x)} =  \arctan{(f_z/f_x)}$. Here the vector magnitude is $|\vec{F}|=F_o\sqrt{f_z^2+f_x^2}=0.1 ~\rm a.u.$. $F_o$ is varied such that $|\vec{F}|$ is constant for the different values of $f_z$ and $f_x$ (see Tables IV-VII).}
\end{figure*}


Here we present results for different force orientations in the $x-z$ plane. The resonance parameters are shown in Fig. \ref{fig:both} for both the $1b_1$ and $3a_1$ orbitals. Due to symmetry the results would be the same for the force in positive $x$ axis as negative $x$ axis, so we choose $ \arctan{(f_z/f_x)}$ to vary in the range $-\pi/2$ to $\pi/2$, where $F_x$ and therefore the $f_x$ are always positive. The lower-case $\{f_x, f_z \}$ are the (dimensionless) fractions  which determine the projection of $\vec{F}$ onto the two axes. These angles corresponds to $\pi$ to $0$ in terms of the polar angle. The two extreme values correspond to forces along $-\hat z$ and $\hat z$ respectively, while the force is in the $\hat{x}$ direction for $\arctan{(f_z/f_x)}=0$. The force strength itself is chosen to be fixed at $|\vec{F}|=0.1$ a.u., not to be confused with $F_o$ which varies throughout the different values of $\arctan{(f_z/f_x)}$.

The orbital resonance positions get closer to one another as the electric force points towards the $z$ axis, and farther as it points towards the $-z$ axis. The $1b_1$ orbital half-width does not reach a peak when $\arctan{(f_z/f_x)}=0$. It reaches its peak around 0.4 radians. The $3a_1$ orbital half-width becomes dominant at larger |arctan| for positive angles than negative. 
\section{Tables}
\subsection{Resonance parameters corresponding to molecular orbitals ($x$ directed force)}
For all the following subsections, the number of significant digits has been controlled at 4 in accord with the findings of Fig. 1 of Ref. \cite{pm21}. This applies foremost to the real part of the energy eigenvalues. The imaginary parts have been controlled to the same number of digits under the \textbf{siunitx} tabular environment. In any case the small imaginary values (below $10^{-7}$) should be trusted to fewer digits. Some of the numbers have been cut off before 4 significant digits by the \textbf{siunitx} tabular environment due to trailing zeroes. The real parts of the energy eigenvalues (Re) are the resonance widths, and the imaginary parts (Im) are the resonance half-widths.

\setlength{\tabcolsep}{6pt}
\bgroup
\def\arraystretch{0.7}
\begin{table}[!h]
\begin{center}
\begin{tabular}{l*{2}{S[round-mode=figures,round-precision=4]}} 

 MO: $1b_1$ & &\\ [0.2ex] 
\hline
\hline
   \rm $\rm F_x$  & $\mathfrak{Re}$ &$\mathfrak{Im}~$ \\ [0.2ex]
   \hline
\hline
   
    0.02 &-0.52139126603452346&-1.2661950169289239E-012\\
    0.04 &-0.52164644366868174&-4.2422749426330813E-007\\
    0.06 &-0.52256820099312140&-1.2518801257310438E-004\\
    0.08&-0.52462855697589139&-1.4748800601492586E-003\\
        0.10&-0.52669611275227401&-5.2131592685208195E-003\\
            0.12&-0.52743343585669622&-1.0973610365391800E-002\\
                0.14&-0.52632493080865450&-1.7972623302145958E-002\\
                    0.16&-0.52318803479703169   & -2.5519664988971343E-002\\
         0.18&-0.51806085489487952      &  -3.2941524239864128E-002\\
0.20 &-0.51122926529868562      & -3.9578185913490042E-002\\
0.22 & -0.50322506975312076      &-4.5022400328886381E-002\\
0.24 &-0.49467515229677400         &-4.9079724971331978E-002
\\
0.26  & -0.48616608085667101  & -5.1950275827738007E-002\\
0.28 & -0.47801208415002339    &  -5.3936134018004969E-002\\
0.30 &-0.47038932512040299        & -5.5403356421138132E-002\\

\hline
\hline

\end{tabular}

\caption{Resonance positions and half-widths for the molecular orbital $1b_1$ of $\rm H_2O$.}
\end{center}
\end{table}
\setlength{\tabcolsep}{6pt}
\bgroup
\def\arraystretch{0.7}
\begin{table}[!h]
\begin{center}
\begin{tabular}{l*{2}{S[round-mode=figures,round-precision=4]}} 

 MO: $3a_1$ & &\\ [0.2ex] 
\hline
\hline
   \rm $\rm F_x$  & $\mathfrak{Re}$ &$\mathfrak{Im}~$ \\ [0.2ex]
   \hline
\hline
   
    0.02 &-0.57227266356679263   & -1.6234131956863717E-012\\
    0.04 &-0.57361397097895728    & -1.6628283786654847E-009\\
    0.06 &-0.57586388848585057  & -2.4239489513153771E-006\\
    0.08&-0.57911724149449750 &-7.9672503363111804E-005\\
        0.10&-0.58342082561199760 &-5.6734603469790128E-004\\
            0.12&-0.58842220781783205   & -1.8500831198222220E-003\\
                0.14&-0.59360620046983958 &-3.9087984055714668E-003\\
                    0.16&-0.59872593526467000  &  -6.4711380645781826E-003\\
         0.18&-0.60377521694281688     &   -9.3451634527189440E-003\\
0.20 &-0.60879597095264137       & -1.2465413063648673E-002\\
0.22 &-0.61379401569588266      & -1.5839317350602670E-002\\
0.24 &-0.61874471841172818        & -1.9462770260082984E-002
\\
0.26  &-0.62362297888015428   &  -2.3338710157779274E-002\\
0.28 &-0.62838866887664158     &  -2.7448867819287039E-002\\
0.30 &-0.63303226768665977         &  -3.1767841057437574E-002\\

\hline
\hline

\end{tabular}

\caption{Resonance positions and half-widths for the molecular orbital $3a_1$ of $\rm H_2O$.}
\end{center}
\end{table}

\setlength{\tabcolsep}{6pt}
\bgroup
\def\arraystretch{0.7}
\begin{table}[!h]
\begin{center}
\begin{tabular}{l*{2}{S[round-mode=figures,round-precision=4]}} 

 MO: $1b_2$ & &\\ [0.2ex] 
\hline
\hline
   \rm $\rm F_x$  & $\mathfrak{Re}$ &$\mathfrak{Im}~$ \\ [0.2ex]
   \hline
\hline
   
    0.02 &-0.71004176036880073      & NA\\
    0.04 &-0.71050073273127123      & NA\\
    0.06 &-0.71127564703686486     & -1.0010701672247280E-010\\
    0.08&-0.71238389124476087   & -6.2741613703632962E-008\\
        0.10&-0.71385761664807523   & -2.5653652290927395E-006\\
            0.12&-0.71575162602411291     &  -2.7491197518352353E-005\\
                0.14&-0.71812820629661200  & -1.3943380828184153E-004\\
                    0.16&-0.72101618869020412     &   -4.4858843406607582E-004\\
         0.18&-0.72438642546272103    &     -1.0756513570173299E-003\\
         0.20&-0.72816241086708788   & -2.1153152754340896E-003\\
0.22 &-0.73224669480262194     &  -3.6219996306833839E-003\\
0.24 &-0.73654262858138708        & -5.6134203137033559E-003\\
0.26 & -0.74096639491749683     &-8.0819766315778068E-003
\\
0.28  &-0.74544926913652132     & -1.1005068889708218E-002\\
0.30 &-0.74993830424685448           &-1.4351652671932790E-002\\

\hline
\hline

\end{tabular}

\caption{Resonance positions and half-widths for the molecular orbital $1b_2$ of $\rm H_2O$.}
\end{center}
\end{table}
\clearpage
\subsection{Dominant resonance parameters vs the force orientation in $x-z$ plane}
\setlength{\tabcolsep}{6pt}
\bgroup
\def\arraystretch{0.7}
\begin{table}[!h]
\begin{center}
\begin{tabular}{l*{3}{S[round-mode=figures,round-precision=4]}} 

 MO: $1b_1$  & &\\ [0.2ex] 
\hline
\hline
   $(f_z,f_x)$&\ $\arctan{(f_z/f_x)}$ & $\mathfrak{Re}~$ &$\mathfrak{Im}~$ \\ [0.2ex]
   \hline
\hline
   
    (0.0, 1.0)& 0 ~rad &-0.52669611276894879    &   -5.2131592591762310E-003\\
    (0.1, 0.9) &0.110657~ rad&-0.52799238107941282     & -5.7659662068946524E-003\\
    (0.2, 0.8) &0.244979 ~rad&-0.52963125447796999     & -6.3673861853992143E-003\\
     (0.3, 0.7) &0.404892 ~rad&-0.53159208805912406     & -6.8818930377488007E-003\\
         (0.4, 0.6) &0.588003 ~rad&-0.53369737396441619     & -6.9954387510055508E-003\\
         (0.5, 0.5) &0.785398 ~rad&-0.53557396240476396     & -6.2741846945050953E-003
\\
         (0.6, 0.4) &0.982794 ~rad&-0.53676670689226280     & -4.6177279359344292E-003
\\
         (0.7, 0.3) &1.1659 ~rad&-0.53711236786233385     & -2.6919133908004217E-003\\
         (0.8, 0.2) &1.32582 ~rad&-0.53699844524314999     & -1.2859499456639330E-003\\
         (0.9, 0.1) &1.46014 ~rad&-0.53684989778788805     & -5.7916433498581411E-004\\
         (1.0, 0.0) &1.57079632679 ~rad&-0.53680151730092707     & -3.9296599652450441E-004\\

\hline
\hline

\end{tabular}
\caption{Resonance positions and half-widths for the molecular orbital $1b_1$ of $\rm H_2O$. The force strengths in each direction are $F_z = F_of_z$ and $F_x = F_o f_x$. Hence the angle of the force from the $x$ to the $z$ axis is $\arctan{(F_of_z/F_of_x)} = \arctan{(f_z/f_x)}$. Here the vector magnitude is $|\vec{F}|=F_o\sqrt{f_z^2+f_x^2}=0.1 ~\rm a.u.$. $F_o$ is varied such that $|\vec{F}|$ is constant.}
\end{center}
\end{table}
\setlength{\tabcolsep}{6pt}
\bgroup
\def\arraystretch{0.7}
\begin{table}[!h]
\begin{center}
\begin{tabular}{l*{3}{S[round-mode=figures,round-precision=4]}} 

 MO: $3a_1$  & &\\ [0.2ex] 
\hline
\hline
   $(f_z,f_x)$&\ $\arctan{(f_z/f_x)}$ & $\mathfrak{Re}~$ &$\mathfrak{Im}~$ \\ [0.2ex]
   \hline
\hline
   
 (0.0, 1.0)& 0 ~rad &-0.58342082567768638&-5.6734604376749043E-004\\
    (0.1, 0.9) &0.110657~ rad&-0.58127873847465150     & -4.0075616358638434E-004\\
    (0.2, 0.8) &0.244979 ~rad&-0.57867770476959723   &   -2.2735886684249079E-004
\\
     (0.3, 0.7) &0.404892 ~rad&-0.57569469307600851     & -1.3620663534365615E-004\\
         (0.4, 0.6) &0.588003 ~rad& -0.57262378657065116     & -3.5327708865819281E-004
\\
         (0.5, 0.5) &0.785398 ~rad&-0.56998443013028122     & -1.2172039792447684E-003\\
         (0.6, 0.4) &0.982794 ~rad&-0.56831660498790293     & -2.8123706301895987E-003\\
         (0.7, 0.3) &1.1659 ~rad&-0.56774501231322716     & -4.5695781831023227E-003\\
         (0.8, 0.2) &1.32582 ~rad&-0.56777594728837033     & -5.8188142605594996E-003\\
         (0.9, 0.1) &1.46014 ~rad& -0.56789999798321711     & -6.4363338979485330E-003
\\
         (1.0, 0.0) &1.57079632679 ~rad&-0.56794362055767977     & -6.5977672391228301E-003
\\

\hline
\hline

\end{tabular}
\caption{Resonance positions and half-widths for the molecular orbital $3a_1$ of $\rm H_2O$. The force strengths in each direction are $F_z = F_of_z$ and $F_x = F_o f_x$. Hence the angle of the force from the $x$ to the $z$ axis is $\arctan{(F_of_z/F_of_x)} = \arctan{(f_z/f_x)}$. Here the vector magnitude is $|\vec{F}|=F_o\sqrt{f_z^2+f_x^2}=0.1 ~\rm a.u.$. $F_o$ is varied such that $|\vec{F}|$ is constant.}
\end{center}
\end{table}
\setlength{\tabcolsep}{6pt}
\bgroup
\def\arraystretch{0.7}
\begin{table}[!h]
\begin{center}
\begin{tabular}{l*{3}{S[round-mode=figures,round-precision=4]}} 

 MO: $1b_1$  & &\\ [0.2ex] 
\hline
\hline
   $(f_z,f_x)$&$\arctan{(f_z/f_x)}$ & $\mathfrak{Re}~$ &$\mathfrak{Im}~$ \\ [0.2ex]
   \hline
\hline
   
    (0.0, 1.0)& 0 ~rad &-0.52669611276894879    &   -5.2131592591762310E-003\\
    (-0.1, 0.9) &-0.110657 ~rad&-0.52547844418590950      & -4.6418154017549909E-003\\
    (-0.2, 0.8) &-0.244979 ~rad&-0.52413395616777958    &  -3.9539735898329520E-003\\
     (-0.3, 0.7) &-0.404892 ~rad&-0.52274993215859389     &  -3.1756477568861889E-003\\
         (-0.4, 0.6) &-0.588003 ~rad&-0.52146505764646101     &  -2.3696744805526268E-003\\
         (-0.5, 0.5) &-0.785398 ~rad&-0.52042044159765233       &  -1.6275477182426748E-003
\\
         (-0.6, 0.4) &-0.982794 ~rad&-0.51968633617771209     &  -1.0341200436836835E-003
\\
         (-0.7, 0.3) &-1.1659 ~rad&-0.51923782533585106       &  -6.2753801634494739E-004\\
         (-0.8, 0.2) &-1.32582 ~rad&-0.51899873420549281      &  -3.9186269444868886E-004\\
         (-0.9, 0.1) &-1.46014 ~rad&-0.51889326559456783      &  -2.8269946867694071E-004\\
         (-1.0, 0.0) &-1.57079632679  ~rad&-0.51886669060851709      &  -2.5463078282055799E-004\\
\hline
\hline
\end{tabular}
\caption{Resonance positions and half-widths for the molecular orbital $1b_1$ of $\rm H_2O$. The force strengths in each direction are $F_z = F_of_z$ and $F_x = F_o f_x$. Hence the angle of the force from the $x$ to the $z$ axis is $\arctan{(F_of_z/F_of_x)} = \arctan{(f_z/f_x)}$. Here the vector magnitude is $|\vec{F}|=F_o\sqrt{f_z^2+f_x^2}=0.1 ~\rm a.u.$. $F_o$ is varied such that $|\vec{F}|$ is constant.}
\end{center}
\end{table}
\setlength{\tabcolsep}{6pt}
\bgroup
\def\arraystretch{0.7}
\begin{table}[!h]
\begin{center}
\begin{tabular}{l*{3}{S[round-mode=figures,round-precision=4]}} 

 MO: $3a_1$  & &\\ [0.2ex] 
\hline
\hline
   $(f_z,f_x)$&$\arctan{(f_z/f_x)}$ & $\mathfrak{Re}~$ &$\mathfrak{Im}~$ \\ [0.2ex]
   \hline
\hline
   
 (0.0, 1.0)& 0 ~rad &-0.58342082567768638&-5.6734604376749043E-004\\
    (-0.1, 0.9) &-0.110657 ~rad&-0.58551685315454916     &   -7.3450174319986421E-004\\
    (-0.2, 0.8) &-0.244979 ~rad&-0.58794954061589033     &     -9.2273202044627542E-004
\\
     (-0.3, 0.7) &-0.404892 ~rad&-0.59062468497220466      &  -1.1149259142925938E-003\\
         (-0.4, 0.6) &-0.588003 ~rad& -0.59332637832469048      & -1.2895706408083967E-003
\\
         (-0.5, 0.5) &-0.785398 ~rad&-0.59575291977652101     &  -1.4293869486065923E-003\\
         (-0.6, 0.4) &-0.982794 ~rad&-0.59764596834174555      &  -1.5281325384200988E-003\\
         (-0.7, 0.3) &-1.1659 ~rad&-0.59891549925369214      & -1.5899126199895855E-003\\
         (-0.8, 0.2) &-1.32582 ~rad&-0.59964019386422984      & -1.6237897328418023E-003\\
         (-0.9, 0.1) &-1.46014 ~rad& -0.59997293499302806      &  -1.6390416679109402E-003
\\
         (-1.0, 0.0) &-1.57079632679  ~rad&-0.60005818360408747      & -1.6429205112015500E-003
\\
\hline
\hline
\end{tabular}
\caption{Resonance positions and half-widths for the molecular orbital $3a_1$ of $\rm H_2O$. The force strengths in each direction are $F_z = F_of_z$ and $F_x = F_o f_x$. Hence the angle of the force from the $x$ to the $z$ axis is $\arctan{(F_of_z/F_of_x)} = \arctan{(f_z/f_x)}$. Here the vector magnitude is $|\vec{F}|=F_o\sqrt{f_z^2+f_x^2}=0.1 ~\rm a.u.$. $F_o$ is varied such that $|\vec{F}|$ is constant.}
\end{center}
\end{table}
\clearpage
\subsection{Resonance parameters corresponding to molecular orbitals ($y$ directed force)}
\setlength{\tabcolsep}{6pt}
\bgroup
\def\arraystretch{0.7}
\begin{table}[!h]
\begin{center}
 \begin{tabular}{l*{2}{S[round-mode=figures,round-precision=4]}} 

 MO: $1b_1$ & &\\ [0.2ex] 
\hline
\hline
   \rm $\rm F_y$  & $\mathfrak{Re}$ &$\mathfrak{Im}$ \\ [0.2ex]
   \hline
\hline
   
    0.02 &-0.52157846302690525    &NA\\
    0.04 &-0.52230656012356969  &   -4.6575412816694994E-010\\
    0.06 &-0.52357355184527599    & -1.1856744822740636E-006\\
    0.08&-0.52550232682856979     &  -4.5318001370007584E-005\\
        0.10&-0.528225407636722658 &  -3.4256067714270779E-004\\
            0.12&-0.53167390457464847   &   -1.2054620665944726E-003\\
                0.14&-0.53560148998332668     &   -2.8247544620603310E-003\\
                    0.16&-0.53974888348419325      &   -5.2270468154104049E-003\\
         0.18&-0.54392158720995987         &    -8.3472216420463277E-003\\
0.20 &-0.54799330742779229         &    -1.2088193578544014E-002\\
0.22 &-0.55188811388078074           &  -1.6352764428416437E-002\\
0.24 &-0.55556166763388959           &  -2.1052184656232896E-002
\\
0.26  & -0.55899279697255355       &   -2.6109842426187139E-002\\
0.28 &-0.56217176110093392     &   -3.1463095623056714E-002\\
0.30 &-0.56509577623930218         &    -3.7056359350214313E-002\\

\hline
\hline

\end{tabular}

\caption{Resonance positions and half-widths for the molecular orbital $1b_1$ of $\rm H_2O$.}
\end{center}
\end{table}
\vspace*{-0.5cm}
\setlength{\tabcolsep}{6pt}
\bgroup
\def\arraystretch{0.7}
\begin{table}[!h]
\begin{center}
 \begin{tabular}{l*{2}{S[round-mode=figures,round-precision=4]}} 

 MO: $3a_1$ & &\\ [0.2ex] 
\hline
\hline
   \rm $\rm F_y$  & $\mathfrak{Re}$ &$\mathfrak{Im}$ \\ [0.2ex]
   \hline
\hline
   
    0.02 &-0.57212751323546718   & NA\\
    0.04 &-0.57304706398786054 &  -2.8399945222645162E-011\\
    0.06 &-0.57462527277592501    & -1.6943798046038278E-007\\
    0.08&-0.57695865660612489    & -1.2432868337150404E-005\\
        0.10&-0.58021249869148528 & -1.4293577635726451E-004\\
            0.12&-0.58448470699271915  &  -6.6255635918048680E-004\\
                0.14&-0.58967788833296908   &  -1.8690081837933186E-003\\
                    0.16&-0.59557302368618503     &   -3.9350172063823980E-003\\
         0.18&-0.60195201015543454        &    -6.8937193573992616E-003\\
0.20 &-0.60865333582142556        &   -1.0687700267489959E-002\\
0.22 &-0.61558493441339424          &  -1.5198190305843098E-002\\
0.24 &-0.62271075555312905           &  -2.0250224136362772E-002
\\
0.26  & -0.63001245037370635   &  -2.5560458255059610E-002\\
0.28 &-0.63735464066635250    &   -3.0658638534732722E-002\\
0.30 &-0.64410916360450188        &   -3.5025179678271887E-002\\

\hline
\hline

\end{tabular}

\caption{Resonance positions and half-widths for the molecular orbital $3a_1$ of $\rm H_2O$.}
\end{center}
\end{table}
\setlength{\tabcolsep}{6pt}
\bgroup
\def\arraystretch{0.7}
\begin{table}[!h]
\begin{center}
 \begin{tabular}{l*{2}{S[round-mode=figures,round-precision=4]}} 

 MO: $1b_2$ & &\\ [0.2ex] 
\hline
\hline
   \rm $\rm F_y$  & $\mathfrak{Re}$ &$\mathfrak{Im}$ \\ [0.2ex]
   \hline
\hline
   
    0.02 &-0.70990801168088846  &-3.1465153255454654E-012\\
    0.04 &-0.70998904928765594 & -5.1000750308569971E-012\\
    0.06 &-0.71022347831891464  &-1.8972615999601483E-007\\
    0.08&-0.71083092410868665  &-2.2183814311186489E-005\\
        0.10&-0.71223856105135508 & -3.2599180828836119E-004\\
            0.12&-0.71469839834213367 & -1.7187394709637677E-003\\
                0.14&-0.71773748798384873 & -5.1385456146262661E-003\\
                    0.16&-0.72039314137192545    &  -1.1012349160739629E-002\\
         0.18&-0.72173077958413601        &   -1.9212325913787937E-002\\
0.20 &-0.72106487645770978       &  -2.9291303215258302E-002\\
0.22 & -0.71798468534997451        & -4.0705236624384536E-002\\
0.24 &-0.71232829380217888          & -5.2920915217166048E-002
\\
0.26  & -0.70415233838140068  & -6.5644326535265030E-002\\
0.28 & -0.69377367940508405   &   -7.8809659512443481E-002\\
0.30 &-0.68215746415658207        &  -9.2701609583246775E-002\\

\hline
\hline

\end{tabular}
\caption{Resonance positions and half-widths for the molecular orbital $1b_2$ of $\rm H_2O$.}
\end{center}
\end{table}

\subsection{Comparison with HF total ionization rates}\label{sec:hfrates}

Table~\ref{tab:totE} contains a direct sums of energy eigenvalues to obtain total Stark shifts and resonance widths, to be compared with with the HF results of Jagau \cite{Jagau2018}. We carry out a direct sum (written $\rm DS ~(2)$ for a sum of the occupied orbitals, where the two indicates double occupation due to spin degeneracy) to find the net energy and shifts. The innermost two orbitals ($1a_1$ and $2a_1$) are left out when their values are too small, which only happens for the widths and not the Stark shifts. The HF row references Jagau \cite{Jagau2018}. \clearpage
\setlength{\tabcolsep}{4pt}
\bgroup
\def\arraystretch{0.7}
\begin{table}[!h]
\begin{center}

 \begin{tabular*} {\columnwidth}{@{\extracolsep{\fill}\extracolsep{0pt}}llllllll}

\hline

 &&$\mathfrak{Re}~\Delta \rm E$&\\ [0.2ex] 
\hline

  $F_x$&&0.04 &  0.06& 0.08&  0.10& 0.12 &0.14\\ [0.2ex]
   \hline

   $1a_1$ &&$-0.000001$&$-0.000002$&$-0.000004$&$-0.000006$&$-0.000009$&$-0.000012$\\
    $2a_1$ &&$-0.000978$&$-0.002202$&$-0.003920$&$-0.006135$&$-0.008853$&$-0.012085$\\
    $1b_2$ & &$-0.000611$&$-0.001386$&$-0.002494$&$-0.003968$&$-0.005862$&$-0.008238$\\
    $3a_1$ &&$-0.001789$&$-0.004039$&$-0.007292$&$-0.011596$&$-0.016597$&$-0.021781$\\
    $1b_1$&&$-0.000306$&$-0.001228$&$-0.003288$&$-0.005356$&$-0.0060933$&$-0.004985$\\
        
         $\rm DS$ (2)&&$-0.007371$&$-0.017715$&$-0.033998$&$-0.054121$&$-0.074829$&$-0.094203$\\
            HF &&$-0.006510$&$-0.015381$&$-0.028639$&$-0.045285$&$-0.063967$&$-0.083259$\\

\hline

 &&$\Gamma$\\ [0.2ex] 
\hline

  $F_x$& &0.04 &  0.06& 0.08&  0.10& 0.12 &0.14\\ [0.2ex]
   \hline

    $1b_2$ &&0.000000&0.000000&0.000000&0.000005&0.000055&0.000279\\
    $3a_1$& &0.000000&0.000005&0.000160&0.001135&0.003700&0.007818\\
    $1b_1$&&0.000000&0.000250&0.002950&0.010426&0.021947&0.035945\\
     $\rm DS ~ (2)$&&0.000002&0.000510&0.006218&0.023132&0.051405&0.088084\\
      
            HF &&0.000004&0.000523&0.004346&0.014036&0.030614&0.052159\\

\hline

 &&$\mathfrak{Re}~\Delta \rm E$&\\ [0.2ex] 
\hline

  $F_y$&&0.04 &  0.06& 0.08&  0.10& 0.12 &0.14\\ [0.2ex]
   \hline

   $1a_1$ &&$-0.000001$&$-0.000002$&$-0.000004$&$-0.000006$&$-0.000009$&$-0.000012$\\
    $2a_1$ &&$-0.001984$&$-0.004459$&$-0.007918$&$-0.012355$&$-0.017767$&$-0.024154$\\
    $1b_2$ & &$-0.000001$&$-0.000334$&$-0.000941$&$-0.002349$&$-0.004809$&$-0.007848$\\
    $3a_1$ &&$-0.001222$&$-0.002800$&$-0.005134$&$-0.008387$&$-0.012660$&$-0.017853$\\
    $1b_1$&&$-0.000966$&$-0.002233$&$-0.004162$&$-0.006885$&$-0.010334$&$-0.014261$\\
        
         $\rm DS$ (2)&&$-0.008545$&$-0.019657$&$-0.036318$&$-0.059966$&$-0.091157$&$-0.128257$\\
            HF &&$-0.007442$&$-0.016968$&$-0.030838$&$-0.049445$&$-0.072722$&$-0.100332$\\

\hline

 &&$\Gamma$\\ [0.2ex] 
\hline

  $F_y$& &0.04 &  0.06& 0.08&  0.10& 0.12 &0.14\\ [0.2ex]
   \hline

    $1b_2$ &&0.000000&0.000000&0.000000&0.000005&0.000055&0.000279\\
    $3a_1$& &0.000000&0.000005&0.000160&0.001135&0.003700&0.007818\\
    $1b_1$&&0.000000&0.000250&0.002950&0.010426&0.021947&0.035945\\
     $\rm DS ~ (2)$&&0.000000&0.000006&0.000220&0.003246&0.014347&0.039327\\
      
            HF &&0.000000&0.000019&0.000387&0.002354&0.007550&0.017486\\

\hline

\end{tabular*}

\caption{\label{tab:totE}
Stark shifts and widths for the valence orbitals of the water molecule for different force strengths. The total molecular Stark shifts and widths calculated using a direct sum of the complex energy eigenvalues of the occupied orbitals, and are labeled DS (2). All results are from the current work except for HF which references Jagau \cite{Jagau2018}.}
\end{center}
\end{table}
\clearpage
\subsection{Tabulations of $\ell$ max = 4 calculations}

\setlength{\tabcolsep}{6pt}
\bgroup
\def\arraystretch{0.7}
\begin{table}[!h]
\begin{center}
 \begin{tabular}{r*{2}{S[round-mode=figures,round-precision=4]}} 

 MO:$~ 1b_1$ & &\\ [0.2ex] 
\hline
\hline
   $\rm  F_z$  & $\mathfrak{Re}$ &$\mathfrak{Im}$ \\ [0.2ex]
   \hline
\hline
       -0.30 & -0.53590465584674674   &  -2.2164978190204972E-002\\
    -0.28 & -0.53413885893409230   &   -1.8781475756867713E-002\\
    -0.26 & -0.53234443242743945 &  -1.5579851602327870E-002 \\
    -0.24 & -0.53051431718286246   & -1.2580032477105220E-002  \\
    -0.22 & -0.52866671024253731    & -9.8245858702223667E-003  \\
    -0.20 & -0.52682296111313942   & -7.3351107809319603E-003 \\
    -0.18 & -0.52500437910446207  &  -5.1584540767034828E-003 \\
    -0.16 & -0.52326190940122896   &  -3.3320500152375591E-003 \\
    -0.14 & -0.52165360211422951  &  -1.8998060048781898E-003 \\
    -0.12 & -0.52027748727399525   & -8.8979871856377448E-004 \\
    -0.10 & -0.51925806582850165  & -2.9965987411586300E-004 \\
    -0.08 & -0.51873298422643177  & -5.4528146033096135E-005 \\
    -0.06 & -0.51878203369155973  &  -2.6687117609046065E-006 \\
    -0.04 &-0.51936914453623595     & NA \\
    -0.02 &-0.52042470474129099   &  NA \\
    0.00 & -0.52192511190368363   & 0  \\
    0.02 & -0.52387750651494414   &  NA\\
    0.04 & -0.52631438114991458    &   -5.9059077162253516E-009\\
    0.06 & -0.52931258385153057   &  -4.2751320958016642E-006 \\
    0.08 & -0.53302668749652626  &  -9.1968630630564991E-005 \\
    0.10 & -0.53755867403659030  &   -5.2554959914654579E-004\\
    0.12 & -0.54279257576784634  &  -1.6053865936564878E-003 \\
    0.14 & -0.54847816115433634  &  -3.4891923937113785E-003 \\
    0.16 & -0.55436813873607504 & -6.1996621098195977E-003 \\
    0.18 & -0.56028029995992890   &  -9.6672721955479712E-003 \\
    0.20 & -0.56608042007141945  &   -1.3812932007653373E-002\\
    0.22 & -0.57169529636103666  &  -1.8523555717147311E-002 \\
    0.24 & -0.57707623410724462    &  -2.3735345637161721E-002 \\
    0.26 & -0.58218134502522012  &  -2.9344103815814749E-002 \\
    0.28 & -0.58703052039612924   &  -3.5295439361277617E-002 \\
    0.30 & -0.59158283134103951  &  -4.1546674666364899E-002 \\

\hline
\hline

\end{tabular}

\caption{Resonance positions and half-widths for the molecular orbital $1b_1$ of $\rm H_2O$.}
\end{center}
\end{table}
\setlength{\tabcolsep}{6pt}
\bgroup
\def\arraystretch{0.7}
\begin{table}[!h]
\begin{center}
 \begin{tabular}{r*{2}{S[round-mode=figures,round-precision=4]}} 

 MO:$~ 3a_1$ & &\\ [0.2ex] 
\hline
\hline
   $\rm  F_z$  & $\mathfrak{Re}$ &$\mathfrak{Im}$ \\ [0.2ex]
   \hline
\hline
    -0.30 &-0.62518346284874926  &-5.4768874110418544E-002\\
    -0.28 &-0.62828573616382544  &-5.0118745004488187E-002\\
    -0.26 &-0.63056591089673497  & -4.4841986064204670E-002\\
    -0.24 &-0.63179238196203469 & -3.9043121697387072E-002\\
    -0.22 &-0.63186539187280655  & -3.2814256814062762E-002\\
    -0.20 &-0.63056180506305426 &-2.6341403816054629E-002\\
    -0.18 &-0.62783605371644968  & -1.9897545634203768E-002\\
    -0.16 &-0.62361610873693329 & -1.3804274499689859E-002\\
    -0.14 &-0.61803472271511883   & -8.4510005683803100E-003\\
    -0.12 &-0.61134301862478013 &-4.2504955508101015E-003\\
    -0.10 &-0.60407034373303781  & -1.5216611421128082E-003\\
    -0.08 &-0.59697703842181915 &-2.8689198284494931E-004\\
    -0.06 &-0.59071145752647924   & -1.3901052193337008E-005\\
    -0.04 &-0.58529718207384007   & -1.9982495249680447E-008\\
    -0.02 &-0.58047214780278744 & NA\\
    0.00 &-0.57616863938358831   &0\\
    0.02 &-0.57246382194350953 & NA\\
    0.04 &-0.56957980565220123   &-8.9459366428701478E-008\\
    0.06 &-0.56809731422368692   & -6.1336640785733011E-005\\
    0.08 &-0.56917068554463179  & -1.2193998599932417E-003\\
    0.10 &-0.57249686430619673 & -6.1491746986767747E-003\\
    0.12 &-0.57546164378034526  & -1.6513935445360364E-002\\
    0.14 &-0.57500196969671413   & -3.2207261093392690E-002\\
    0.16 &-0.56783992138424355 &-5.1951578714038528E-002\\
    0.18 &-0.55033925185196220  &-7.0718838384650082E-002\\
    0.20 &-0.52592627852209584 & -7.9213706484822677E-002\\
    0.22 &-0.50556136207793789  &-7.9484338359527090E-002\\
    0.24 &-0.48956574680852621  & -7.7291969849304656E-002\\
    0.26 &-0.47646762307122348   &-7.4975193585312946E-002\\
    0.28 &-0.46510081292690292  & -7.2799067402271336E-002\\
    0.30 &-0.45503127030914614  &-7.1041427979376237E-002\\

\hline
\hline

\end{tabular}

\caption{Resonance positions and half-widths for the molecular orbital $3a_1$ of $\rm H_2O$.}
\end{center}
\end{table}
\setlength{\tabcolsep}{6pt}
\bgroup
\def\arraystretch{0.7}
\begin{table}[!h]
\begin{center}
 \begin{tabular}{r*{2}{S[round-mode=figures,round-precision=4]}} 

 MO:$~ 1b_2$ & &\\ [0.2ex] 
\hline
\hline
   $\rm  F_z$  & $\mathfrak{Re}$ &$\mathfrak{Im}$ \\ [0.2ex]
   \hline
\hline
    -0.30 &-0.67629125367865961  & -9.0215051209886505E-003\\
    -0.28 &-0.67679170671996092 &-7.0068720978143867E-003\\
    -0.26 &-0.67745863029276876  &-5.2395462274503549E-003\\
    -0.24 &-0.67833030242850834  &-3.7353284478876098E-003\\
    -0.22 &-0.67944639657083283 & -2.4998273259298855E-003\\
    -0.20 &-0.68085514573347083  &-1.5394576632266432E-003\\
    -0.18 &-0.68261274093233104  & -8.4460057000921879E-004\\
    -0.16 &-0.68477684846316256 &-3.9385511474606408E-004\\
    -0.14 &-0.68740184338799992  & -1.4418485553340435E-004\\
    -0.12 &-0.69052503078590333  & -3.6315309926249236E-005\\
    -0.10 &-0.69415614387600200  & -4.9141753395191015E-006\\
    -0.08 &-0.69827913239129669   & -2.1531866324888371E-007\\
    -0.06 &-0.70287167449953114 & -9.1547595406675198E-010\\
    -0.04 &-0.70792391538981014 &NA\\
    -0.02 &-0.71343903060596203  & NA\\
    0.00 &-0.71942859363488965 &0\\
    0.02 &-0.72591115747023771  &NA\\
    0.04 &-0.73291269095895417  & NA\\
    0.06 &-0.74046851498353283 & -1.2754536525154512E-009\\
    0.08 &-0.74862836960553591 &-3.3973220543349827E-007\\
    0.10 &-0.75746876033865973 & -8.6666474102590980E-006\\
    0.12 &-0.76709928412753881  &-7.0478387133456719E-005\\
    0.14 &-0.77762762687059828 & -3.0278303720788818E-004\\
    0.16 &-0.78910325679319071 & -8.8212158127819228E-004\\
    0.18 &-0.80149541699828220  &-1.9916796981336300E-003\\
    0.20 &-0.81470674530043541  &-3.7811456322276982E-003\\
    0.22 &-0.82861494750904252  & -6.3412672051028516E-003\\
    0.24 &-0.84308957316524702    &-9.7238013360587544E-003\\
    0.26 &-0.85801480040491007  & -1.3927155021298071E-002\\
    0.28 &-0.87330490980112407 & -1.8945677567944474E-002\\
    0.30 &-0.88886704744584810  & -2.4751587543660511E-002\\

\hline
\hline

\end{tabular}

\caption{Resonance positions and half-widths for the molecular orbital $1b_2$ of $\rm H_2O$.}
\end{center}
\end{table}
\clearpage
\bibliography{Partial2}